\documentclass[12pt]{article} 
\usepackage{latexsym,amssymb} 
\newcommand{\ft}[2]{{\textstyle\frac{#1}{#2}}} 

\def\1bar{1\hskip -.275cm -} 
\def\2bar{2\hskip -.275cm -} 
\def\3bar{3\hskip -.275cm -} 
\newcommand{\size}{\tiny} 
\begin{document} 
\begin{titlepage} 
\begin{flushright} 
Preprint DFTT 49/99\\ 
hep-th/9909188 \\ 
September 1999\\ 
\end{flushright} 
\vskip 2cm 
\begin{center} 
{{\Large \bf 
The structure of  $\mathcal{N}=3$ multiplets in $AdS_4$ 
 and the complete $Osp(3 \vert 4)\times SU(3)$ spectrum 
of M-theory on $AdS_4 \times N^{0,1,0}$. \hskip 0.2cm $^\dagger$ 
}}\\ 
\vfill 
{\large Pietro Fr\'e, Leonardo Gualtieri  and Piet Termonia} \\ 
\vfill 
{ \sl  Dipartimento di Fisica Teorica, Universit\'a di Torino, via P. 
Giuria 1, 
I-10125 Torino, \\ 
 Istituto Nazionale di Fisica Nucleare (INFN) - Sezione di Torino, 
Italy \\} 
\end{center} 
\vfill 
\begin{abstract} 
In this paper, relying on previous results of one of us on harmonic 
analysis, we derive the complete spectrum of $Osp(3\vert 4) \times SU(3)$ 
multiplets that one obtains compactifying D=11 supergravity on the 
unique homogeneous space $N^{0,1,0}$ that has a tri-sasakian 
structure, namely leads to $\mathcal{N}=3$ supersymmetry both in the 
four--dimensional bulk and on the three--dimensional boundary.  As in 
previously analyzed cases the knowledge of the Kaluza Klein spectrum, 
together with general information on the geometric structure of the 
compact manifold is an essential ingredient to guess and construct 
the corresponding superconformal field theory. This is work in 
progress. As a bonus of our analysis we derive and present the 
explicit structure of all unitary irreducible representations of the 
superalgebra $Osp(3\vert 4)$ with maximal spin content $s_{max} \le 
2$. 
\end{abstract} 
\vspace{2mm} \vfill \hrule width 3.cm 
{\footnotesize 
 $^ \dagger $ \hskip 0.1cm Supported by   EEC  under TMR contract 
 ERBFMRX-CT96-0045} 
\end{titlepage} 
\section{Introduction} 
In the last two years, since the first proposals by J. Maldacena \cite{maldapasto} 
and Witten \cite{witten}, 
the studies on the $AdS/CFT$ correspondence between 
superconformal field theories on the boundary $\partial AdS_{p+2}$ of 
$p+2$--dimensional anti de Sitter space and bulk supergravity on 
$AdS_{p+2} \times X^{D-p-2}$ were developed in an extraordinarily large 
literature. For an overview of the field, we just refer to the extensive 
review \cite{maldaetl} and to all references therein. 
In the beginning attention has mainly concentrated on the case $p=3$, 
$D=10$ where one can gain non trivial information on the  infrared 
conformal fixed point of $d=4$ supersymmetric gauge theories 
from $D3$--branes and Kaluza Klein compactifications 
of type IIB supergravity on suitable compact $5$--manifolds $X^5$. 
Besides the case $X^5=S^7$ that leads to $N=4$ Yang Mills on the boundary, 
there is just one instance of homogeneous $5$--manifold, namely 
$X^5=T^{1,1}$, that leads to $N=1$ Yang Mills. For this case, 
 where the superconformal field theory 
was proposed by  Klebanov and Witten \cite{witkleb}, a complete 
comparison with the Kaluza Klein analysis was given in the spring of this year by 
Ceresole, D'Auria, Dall'Agata and Ferrara in \cite{sergiotorino}. 
Since the beginning of this year, however, 
  interest has developed on the $p=2$, $D=11$ case where 
information can be  gained on three--dimensional conformal field theories 
from $M2$--branes and Kaluza Klein compactifications of $D=11$ 
supergravity on $AdS_4 \times X^7$. In this context one can  rely on 
a wealth of results built up in the early eighties when Kaluza Klein 
supergravity was very topical in a totally different perspective (see 
\cite{duffrev} for an overview). In  particular as a very significant heritage from 
that scientific season we have: 
{\sl a)} {the complete classification of compact homogeneous 
  7-manifolds $G/H$ that can be used as internal dimensions, together 
  with their number of unbroken supersymmetries $N(G/H)$ 
  \cite{castromwar},} 
  {\sl b)}{the development of systematic algebraic techniques in 
  harmonic analysis \cite{casher} 
  \cite{spectfer},\cite{univer},\cite{spec321},\cite{castnpqr} 
  to derive the Kaluza Klein mass--spectra \cite{bosmass} ,} 
  {\sl c)} { the algebraic technology 
  \cite{gunawar,frenico,multanna} to construct and classify the 
  unitary irreducible representations ($UIR$.s) of the $Osp(N\vert 
  4)$ superalgebras, whose role in Kaluza Klein supergravity was 
  first pointed out in \cite{osp48}}. 
\par 
Although large, the heritage of the eighties is incomplete in two 
respects: {\it 1)} the structure of {\it all} long and short $Osp(N\vert 4)$ 
multiplets relevant to Kaluza Klein mass spectra \footnote{The 
supermultiplets relevant to Kaluza Klein theories are those with 
maximal spin state $s_{max} \le 2$} was not entirely determined, {\it 
2)} the harmonic analysis and the spectra of all supersymmetric $G/H$ 
7--manifolds had not been calculated in an exhaustive way, 
only partial results existing in that literature. 
\cite{spectfer,spec321,pagepopeM,pagepopeQ}. In the last year a 
programme has been  carried through 
\cite{m111,susp,adscftcheckers,n010massspectrum,n3conf,noiq111,v52} 
that aims at completing the Kaluza Klein spectra of all 
supersymmetric homogeneous $7$--manifolds 
and deriving the corresponding superconformal field theories in 
$D=3$. In \cite{m111} the $KK$ spectrum of the manifold $M^{1,1,1}$ 
was determined and those results of harmonic analysis were also used to 
derive the complete structure of $Osp(2\vert 4)$ supermultiplets. 
This made it possible to perform a general study of the 
$3D$ superconformal field theories associated with sasakian 
$7$--manifolds and make a detailed comparison between Kaluza Klein 
spectra and the spectra of composite conformal operators on the 
boundary. Indeed the general geometric framework for the formulation 
of such three--dimensional SCFT.s was lead down in \cite{adscftcheckers} 
where the $M^{1,1,1}$ and $Q^{1,1,1}$ cases were explicitly 
constructed. In the first case the comparison with Kaluza Klein 
results was already available and showed perfect agreement together 
with new predictions. The comparison for the $Q^{1,1,1}$ case has 
been done in a separate publication which is to appear 
\cite{noiq111}. Also here perfect agreement has been found; furthermore the 
analysis of the $Q^{1,1,1}$ theory has been extended in \cite{korean1} 
to include also domain walls. 
\par 
In the list of supersymmetric $7$--manifolds of \cite{castromwar} 
there is a unique case with $8> \mathcal{N} > 2$, namely the 
$\mathcal{N}=3$ manifold $N^{0,1,0}\equiv SU(3)\times SU(2)/SU(2) \times 
U(1)$. This is a particularly interesting case where the 
sasakian $7$--manifold is actually tri-sasakian, the necessary 
geometric basis to realize $\mathcal{N}=3$ supersymmetry. Furthermore 
since in $d=3$ the $\mathcal{N}=3$ gauge and matter multiplets are 
automatically $\mathcal{N}=4$ multiplets the SCFT associated with 
$N^{0,1,0}$ must be realized as a suitable deformation of an 
$\mathcal{N}=4$ gauge theory which makes this unique case very 
interesting also in view of its very simple flavor group 
$G_{flavor}=SU(3)$. This was the motivation for a systematic 
completion of the harmonic analysis also on this manifold, which was 
accomplished in \cite{n010massspectrum}. 
\par 
In the present letter we use the results of \cite{n010massspectrum} to 
obtain two results: 
\begin{enumerate} 
  \item The general structure of long and short $Osp(3\vert 4)$ supermultiplets with 
  $s_{max} \le 2$ together with their decomposition into $Osp(2\vert4)$ 
  supermultiplets. 
  \item The specific $SU(3)$ representation assignment of the 
  $Osp(3\vert 4)$ supermultiplets actually appearing in the Kaluza 
  Klein spectrum on $N^{0,1,0}$. 
\end{enumerate} 
The first point in the above list has a general algebraic validity 
and aims at the completion of the missing results on $UIR$.s of the 
$Osp( \mathcal{N}\vert 4)$ superalgebra. The second point, instead, 
is instrumental to the construction of the $\mathcal{N}=3$ SCFT 
associated with $N^{0,1,0}$ which is presently under study 
\cite{n3conf}. 
\section{$Osp(3\vert 4)$ multiplets} 
Let us briefly  explain how the structure of the $Osp\left(3\vert 4\right)$ 
multiplets has been obtained. 
We have determined them 
by means of the following tools: 
\begin{itemize} 
\item the method of induced representations, as in \cite{frenico,multanna} 
\item the harmonic analysis of $N^{0,1,0}$  \cite{n010massspectrum}, 
\item the ${\cal N}=3\longrightarrow{\cal N}=2$ decomposition. 
\end{itemize} 
The first tool gives the structure of long multiplets, the 
other two give the short multiplet structures. 
Representations of $Osp\left(3\vert 4\right)$ are labeled by the 
representation of the maximal compact subgroup $SO(2)_E \times 
SO(3)_{spin} \times SO(3)_R \subset Osp(3\vert 4)$ to which the 
Clifford vacuum $| E_0, s_0 , J_0 >$ is assigned. As anticipated by 
our notation this means to specify the energy $E_0 \in \mathbb{R}$, the 
spin $s_0\in\mathbb{Z}/2 $ and 
the $R-$symmetry representation  $J_0\in\mathbb{Z}/2$ 
which we call isospin. 
The representations are unitary, provided 
\begin{eqnarray} 
E_0&\ge& J_0+s_0+1\\ 
{\rm or}&&\nonumber\\ 
E_0&=&J_0,\,~~~s_0=0. 
\end{eqnarray} 
If the unitarity bound is saturated, some states have vanishing norms 
and decouple, so we have a shortened representation. 
These shortened representations are particularly interesting from 
the viewpoint of the $AdS/CFT$ correspondence since they are $BPS$ states, 
dual to primary conformal operators of the boundary SCFT. 
These states are annihilated by some combination of the 
$Osp\left(3\vert 4\right)$ supercharges 
and the dual boundary operators satisfy differential superfield 
conditions \cite{susp}. 
\\ 
We have short representations when 
\begin{equation} 
E_0=J_0+s_0+1~~{\rm or}~E_0=J_0,~s_0=0. 
\end{equation} 
We stress that there is another  shortening mechanism of 
a completely different origin. The creation operators that act on the 
Clifford vacuum have isospin $0\le J_0\le 2$. If the isospin of the Clifford vacuum 
is $J_0\ge 2$, the creation operators give rise to states with isospin 
in the range 
$J_0-J\le J_{\rm composite}\le J_0+J$. Yet, in the case where $J_0=0,1$, 
some of these states cannot appear. This mechanism is not related 
to an unitarity bound, and these representations are not $BPS$ stated of 
supergravity, nor primary conformal operators on the boundary. Then we 
call {\it long} the representations with $E_0>s_0+J_0+1$, even if $J_0=0,1$. 
In this context, the massless representations are the short ones with $J_0=0$ 
for the case of the massless graviton and gravitino multiplets and with $J_0=1$ for 
the case of the massless vector multiplets. 
\\ 
For convenience we adopt the following notations. We denote 
\begin{equation} 
SD\left(s_{\rm max},E_0,J_0\vert 3\right) 
\end{equation} 
an UIR of $Osp\left(3\vert 4\right)$ with maximal spin $s_{\rm max}$, 
Clifford vacuum energy and isospin $E_0,~J_0$. Note that, differently from \cite{frenico}, 
we use $s_{\rm max}$ instead of the Clifford vacuum spin $s_0$. 
This notation can be extended to all $Osp\left({\cal N}\vert 4\right)$ 
representations: for example, we denote the ${\cal N}=2$ UIRs by 
\begin{equation} 
SD\left(s_{max},E_0,y_0\vert 2\right), 
\end{equation} 
where $y_0$ is the Clifford vacuum hypercharge. 
\\ 
The complete list of the $Osp\left(3\vert 4\right)$ UIRs with $s_{max}\le 2$ 
is given below: 
\begin{itemize} 
\item {\em long graviton multiplet} $SD\left(2,E_0,J_0\vert 3\right)$ where $E_0>J_0+{3/2}$, see 
table \ref{longgraviton}; 
\item {\em long gravitino multiplet} $SD\left(3/2,E_0,J_0\vert 3\right)$ where $E_0>J_0+1$, see 
table \ref{longgravitino}; 
\item {\em short graviton multiplet} $SD\left(2,J_0+{3/ 2},J_0\vert 3\right)$, see 
table \ref{shortgraviton}; 
\item {\em short gravitino multiplet} $SD\left(3/2,J_0+1,J_0\vert 3\right)$, see 
table \ref{shortgravitino}; 
\item {\em short vector multiplet} $SD\left(1,J_0,J_0\vert 3\right)$, see 
table \ref{shortvector}. 
\end{itemize} 
Note that there are no long vector multiplets, and no hypermultiplets at all. 
\\ 
The ${\cal N}=3\longrightarrow{\cal N}=2$ decompositions of 
the above multiplets, which is essential for 
the construction of the dual superconformal field theory on the 
$AdS$--boundary \cite{n3conf}, are listed below: 
\begin{eqnarray} 
SD\left(2,E_0,J_0\vert 3\right)&\longrightarrow&\bigoplus\limits_{y=-J_0}^{J_0}SD 
\left(2,E_0+1/2,y\vert 2\right) 
\oplus\bigoplus\limits_{y=-J_0}^{J_0}SD\left(3/2,E_0,y\vert 2\right) 
\nonumber\\ 
&&\oplus\bigoplus\limits_{y=-J_0}^{J_0}SD\left(3/2,E_0+1,y\vert 2\right)\oplus 
\bigoplus\limits_{y=-J_0}^{J_0}SD\left(1,E_0+1/2,y\vert 2\right) 
\nonumber\\ 
&&{\rm where~}E_0>J_0+3/2
\nonumber\\
SD\left(2,J_0+3/2,J_0\vert 3\right)&\longrightarrow&\bigoplus 
\limits_{y=-J_0}^{J_0}SD\left(2,J_0+2,y\vert 2\right) 
\oplus\bigoplus\limits_{y=-J_0}^{J_0}SD\left(3/2,J_0+3/2,y\vert 2\right) 
\nonumber\\
SD\left(3/2,E_0,J_0\vert 3\right)&\longrightarrow&\bigoplus 
\limits_{y=-J_0}^{J_0}SD\left(3/2,E_0+1/2,y\vert 2\right) 
\oplus\bigoplus\limits_{y=-J_0}^{J_0}SD\left(1,E_0+1,y\vert 2\right)
\nonumber\\ 
&&\oplus\bigoplus\limits_{y=-J_0}^{J_0}SD\left(1,E_0,y\vert 2\right)~~~~~ 
{\rm where~}E_0>J_0+1
\nonumber\\
SD\left(3/2,J_0+1,J_0\vert 3\right)&\longrightarrow&\bigoplus 
\limits_{y=-J_0}^{J_0}SD\left(3/2,J_0+3/2,y\vert 2\right) 
\oplus\bigoplus\limits_{y=-J_0}^{J_0}SD\left(1,J_0+1,y\vert 2\right) 
\nonumber\\
SD\left(1,J_0,J_0\vert 3\right)&\longrightarrow&\bigoplus 
\limits_{y=-J_0+1}^{J_0-1}SD\left(1,J_0,y\vert 2\right) 
\oplus SD\left(1/2,J_0,J_0\right\vert 2)\nonumber\\ 
&&\oplus SD\left(1/2,J_0,-J_0\right\vert 2). 
\label{decomposition}
\end{eqnarray} 
For the case $\mathcal{N}=2$ the $Osp\left(2\vert 4\right)$ UIRs with $s_{max}\le 2$ 
were determined in \cite{m111} relying on harmonic analysis on the manifold 
$M^{1,1,1}$ plus previous partial results \cite{frenico,multanna,gunawar}.
They are 
\footnote{Notice that the massless ${\cal N}= 
2$ graviton, gravitino and vector multiplets 
are the short ones with $y_0=0$.}: 
\begin{itemize} 
\item {\em long graviton multiplet} $SD\left(2,E_0,y_0\vert 2\right)$ where $E_0>y_0+2$, 
\item {\em long gravitino multiplet} $SD\left(3/2,E_0,y_0\vert 2\right)$ where $E_0>y_0+3/2$ 
\item {\em long vector multiplet} $SD\left(1,E_0,y_0\vert 2\right)$ where $E_0>y_0+1$ 
\item {\em short graviton multiplet} $SD\left(2,y_0+2,y_0\vert 2\right)$, 
\item {\em short gravitino multiplet} $SD\left(3/2,y_0+3/2,y_0\vert 2\right)$ 
\item {\em short vector multiplet} $SD\left(1,y_0+1,y_0\vert 2\right)$, 
\item {\em hypermultiplet} $SD\left(1/2,y_0,y_0\vert 2\right)\oplus SD\left(1/2,y_0,-y_0\vert 2\right)$. 
\end{itemize} 
 
\section{The $N^{0,1,0}$ spectrum} 
The multiplets presented in this paper  are 
all the $Osp(3\vert 4)$ multiplets that appear compactifying 
$D=11$ supergravity on $AdS_4 \times N^{0,1,0}$. They are obtained 
by using the Kaluza Klein mass spectrum derived in \cite{n010massspectrum} 
\footnote{We refer to the original papers \cite{castromwar}, 
\cite{castn010}, \cite{castnpqr} and also to \cite{n010massspectrum} 
for details on the geometry of the manifold $N^{0,1,0}$} 
together with the mass relations in \cite{univer,noi321} and 
the multiplet structure of the long ${\cal N}=3$ 
multiplets. 
\par 
The procedure that we used is exactly the 
same as that employed in \cite{m111}. We repeat here the 
essential steps. In \cite{n010massspectrum} 
the masses were calculated of the spin-$2$ field 
$h$, of all the spin-$\ft32$ fields $\chi^+, \chi^-$ 
and of all the spin $1$-fields $A, W, Z$. 
Then for each spin-$2$ there is one graviton multiplet. 
We fill the rest of the multiplet with the 
spin-$\ft32$ and spin-$1$ fields. The 
spin-$\ft32$ fields that are left over are the highest spin 
components of the gravitino multiplets. We also 
fill them out with the spin-$1$ fields. The spin-$1$ 
that are finally left over are the highest component fields 
of the vector multiplets. These latter are necessarily short since 
there is no long vector multiplets for $Osp(3\vert 4)$ as we have 
already explained. 
\par 
This procedure gives the upper-spin parts ({\it i.e.} $s \ge 1$) of the 
$Osp(3\vert 4)$ multiplets. In 
\cite{n010massspectrum} also the masses of the fields 
$\lambda_L, S, \Sigma$ were calculated.  These are the spin $1/2$ fields 
coming from the longitudinal part of the $D=11$ gravitino $\Psi_\mu$, while 
$S$ and $\Sigma$ are spin zero fields coming from the trace part of internal 
metric $g_{ij}$\footnote{The Kaluza Klein notations we use are those of 
\cite{m111} which were established long ago in \cite{spectfer,univer} 
and are anyhow reviewed in the book \cite{castdauriafre}}. With this information 
we can also identify some of the spin zero and spin one half states 
occurring in the already predicted multiplets, which provides an important consistency 
check on the whole deduction. 
\par 
Summarizing, the conclusion is that the information on the upper-spin components 
that we have from \cite{n010massspectrum}  is sufficient (apart from consistency checks) 
to fill out all the multiplets. One just uses supersymmetry. 
Instead of using ${\cal N}=3$ supersymmetry, 
it is however more convenient \cite{adscftcheckers} 
to consider the $Osp(3\vert 4)$ multiplets  decomposed into 
 $Osp(2\vert 4)$ multiplets (\ref{decomposition}) 
whose structure was derived 
in \cite{m111} and whose description as constrained superfields is known 
from \cite{susp}. 
We use the notation of \cite{n010massspectrum}, where 
\begin{equation} 
H_0={16\over 3}\left(2\left(M_1^2+M_2^2+M_1M_2+3M_1+3M_2\right)-3J\left(J+1\right) 
\right). 
\end{equation} 
and $(M_1,M_2,J)$ are the labels of an irreducible $SU(3) \times 
SU(2)$ representation according to: 
\begin{eqnarray*} 
\begin{array}{l} 
\begin{array}{|c|c|c|c|c|c|} 
\hline 
             \hskip .3 cm & \cdots & \hskip .3 cm & 
             \hskip .3 cm & \cdots & \hskip .3 cm \\ 
\hline 
\end{array}\,,\\ 
\begin{array}{|c|c|c|} 
             \hskip .3 cm & \cdots & \hskip .3 cm \\ 
\hline 
\end{array} 
\end{array}\\ 
\underbrace{\hskip 2.2 cm}_{M_2} 
\underbrace{\hskip 2.2 cm}_{M_1} 
\end{eqnarray*} 
for $SU(3)$ and 
\begin{eqnarray*} 
\begin{array}{|c|c|c|} 
\hline 
             \hskip .3 cm & \cdots & \hskip .3 cm \\ 
\hline 
\end{array}\,.\\ 
\underbrace{\hskip 2.2 cm}_{2J} 
\end{eqnarray*} 
for $SU(2)$. 
\subsection{Long multiplets} 
There are long multiplets for the following $SU\left(3\right)$ representations: 
\begin{equation} 
\cases{M_1=k~~~~~~~~k\ge 0\cr M_2=k+3j~~j\ge 0\cr} 
\end{equation} 
$k,j$ integers. 
\begin{itemize} 
\item For every $SU\left(3\right)$ representation with $k\ge 0,~j\ge 2$ there is only 
one of the following multiplets, that are long: 
\begin{equation} 
\begin{array}{|c|c|c|} 
\hline 
{\rm multiplet} & {isospin} & {\rm energy} \\ 
\hline 
SD\left(E_0,2,J_0\right) & j\le J_0\le k+j & E_0={1\over 4}\sqrt{H_0+36} \\ 
\hline 
SD\left(E_0,3/2,J_0\right) & j\le J_0\le k+j & E_0={1\over 4}\sqrt{H_0+36}-{3\over 2} \\ 
\hline 
SD\left(E_0,3/2,J_0\right) & j\le J_0\le k+j & E_0={1\over 4}\sqrt{H_0+36}+{3\over 2} \\ 
\hline 
\end{array} 
\end{equation} 
\item For every $SU\left(3\right)$ representation with $k\ge 0,~j=1$ there is only 
one of the following multiplets, that are long: 
\begin{equation} 
\begin{array}{|c|c|c|} 
\hline 
{\rm multiplet} & {isospin} & {\rm energy} \\ 
\hline 
SD\left(E_0,2,J_0\right) & 1\le J_0\le k+1 & E_0={1\over 4}\sqrt{H_0+36} \\ 
\hline 
SD\left(E_0,3/2,J_0\right) & 1\le J_0< k+1 & E_0={1\over 4}\sqrt{H_0+36}-{3\over 2} \\ 
\hline 
SD\left(E_0,3/2,J_0\right) &1\le J_0\le k+1 & E_0={1\over 4}\sqrt{H_0+36}+{3\over 2} \\ 
\hline 
\end{array} 
\end{equation} 
\item For every $SU\left(3\right)$ representation with $k\ge 0,~j=0$ there is only 
one of the following multiplets, that are long: 
\begin{equation} 
\begin{array}{|c|c|c|} 
\hline 
{\rm multiplet} & {isospin} & {\rm energy} \\ 
\hline 
SD\left(E_0,2,J_0\right) & 0\le J_0< k & E_0={1\over 4}\sqrt{H_0+36} \\ 
\hline 
SD\left(E_0,3/2,J_0\right) & 0\le J_0< k & E_0={1\over 4}\sqrt{H_0+36}-{3\over 2} \\ 
\hline 
SD\left(E_0,3/2,J_0\right) &0\le J_0\le k & E_0={1\over 4}\sqrt{H_0+36}+{3\over 2} \\ 
\hline 
\end{array} 
\end{equation} 
\end{itemize} 
\subsection{Short multiplets} 
There are the following short multiplets in the following $SU\left(3\right)$ 
 representations: 
\begin{itemize} 
\item There is only one massive short graviton multiplet $SD\left(J_0+3/2,2,J_0\right)$ 
in each of the representations: 
\begin{equation} 
M_1=k,~~M_2=k,~~k\ge 1. 
\end{equation} 
It has 
\begin{equation} 
E_0=k+3/2,~J_0=k. 
\end{equation} 
\item There is only one massive short gravitino multiplet $SD\left(J_0+1,3/2,J_0\right)$ 
in each of the representations: 
\begin{equation} 
M_1=k,~~M_2=k+3,~~k\ge 0. 
\end{equation} 
It has 
\begin{equation} 
E_0=k+2,~J_0=k+1. 
\end{equation} 
\item There is only one massive short vector multiplet $SD\left(J_0,1,J_0\right)$ 
in each of the representations: 
\begin{equation} 
M_1=k,~~M_2=k,~~k\ge 2. 
\end{equation} 
It has 
\begin{equation} 
E_0=k,~J_0=k. 
\end{equation} 
\end{itemize} 
\subsection{Massless multiplets} 
The massless sector of the theory is composed by the following multiplets.
\begin{itemize} 
\item There is  one massless graviton multiplet in the representation: 
\begin{equation} 
M_1=M_2=0 
\end{equation} 
It has 
\begin{equation} 
E_0=3/2,~J_0=0. 
\end{equation} 
This multiplet  has the standard field content expected for the 
$\mathcal{N}=3$ supergravity multiplet in four--dimensions, 
namely  one massless graviton, three massless gravitinos 
that gauge    ${\cal N}=3$ supersymmetry,  three 
massless vector fields (organized in a $J_0=1$ adjoint representation of $SO(3)_R$) 
that gauge the  $R$-symmetry and one spin one--half field. 
\item There is one massless vector multiplet in each of the representations: 
\begin{eqnarray} 
M_1=M_2=1\label{killing}\\ 
M_1=M_2=0\label{betti}\,.
\end{eqnarray} 
They have: 
\begin{equation} 
E_0=1,~J_0=1. 
\end{equation} 
The multiplet (\ref{killing}) contains the gauge vectors of the 
$SU\left(3\right)$ isometry. 
The multiplet (\ref{betti}) is the Betti multiplet \cite{spec321}, 
related to the non--trivial cohomology of $N^{0,1,0}$ in degree two. 
The fact that there is just one Betti multiplet is in perfect 
agreement with the results of \cite{adscftcheckers} where it was 
shown that $H^2\left(N^{0,1,0}\right)=\mathbb{Z} $. All the Kaluza Klein 
states are neutral with respect to the Betti multiplet, but according to the 
analysis of \cite{adscftcheckers} we expect that the SCFT should 
contain non--perturbative {\it baryon states} carrying Betti charges 
which can be reinterpreted as $5$--branes wrapped on supersymmetric 
$5$--cycles of $N^{0,1,0}$. 
\end{itemize} 
\section{Conclusive Remarks} 
From the present algebraic analysis we conclude that, seen as an 
$\mathcal{N}=2$ theory obtained from a sasakian $7$--manifold, the superconformal field 
theory of $N^{0,1,0}$ contains the following hypermultiplets\footnote{Note that the 
hypermultiplets with $k\ge 2$ come from the decomposition of the $ \mathcal{N}=3$ short vector 
multiplets, while the hypermultiplet with $k=1$ arises from the decomposition of 
the $ \mathcal{N}=3$ massless vector multiplet}: 
\begin{equation} 
  M_1=M_2=E_0=y_0=k \quad ;\quad k\ge 1. 
\label{chiralring} 
\end{equation} 
which constitute the chiral ring of the model. Following the 
ideas of \cite{adscftcheckers} it is now a geometric challenge for us 
to construct, starting from a suitable description of $N^{0,1,0}$ a 
three--dimensional gauge theory that reproduces this chiral ring and 
the rest of the Kaluza Klein spectrum \cite{n3conf}. We also note 
that for the first time in the history of the $AdS/CFT$ 
correspondence the dual bulk supergravity is a gauged $\mathcal{N}=3$ 
Lagrangian. As derived and explained in \cite{noinis3} 
such   lagrangians display a unique scalar geometry $SU(3,n)/SU(3)\times SU(n)\times U(1)$ 
and are fixed by giving the 
number of massless vector multiplets $n$ and the gauge group structure 
constants. In our case the number of vector multiplets is $n=8+1=9$ so 
that the scalar manifold is $SU(3,9)/SU(3)\times SU(9)\times U(1) $ 
while the gauge group is $SO(3)_R\times SU(3)$ 
\vskip 0.3cm 
\leftline{\bf Acknowledgements} The authors are grateful to their collaborators 
M. Bill\'o, A. Zaffaroni,  D. Fabbri, C. Reina and P. Merlatti for many important and 
clarifying discussions.  One of us (P. Fr\'e) would also like to 
thank his old time friend R. D'Auria for the continuous and fruitful 
exchange of information on the development of parallel projects 
at the  University of Torino and at the Politecnico of Torino.

\begin{table}[htbp] 
  \begin{center} 
    \size 
    \begin{tabular}{|c|c|c|} 
      \hline 
      \multicolumn{3}{|c|}{} \cr 
      \multicolumn{3}{|c|}{$DS(2,E_0>J_0+3/2,J_0\ge 2\vert 3)$} \cr 
      \multicolumn{3}{|c|}{} \cr 
      \hline 
      spin & energy & isospin  \cr 
      \hline 
      \hline 
       &&\cr 
      $2$ & $E_0+\ft32$ & $J_0$ \cr 
       &&\cr 
      \hline 
       &&\cr 
        & $E_0+2$ & $\cases{J_0+1 \cr J_0 \cr J_0-1}$ 
      \cr 
      $\ft32$ & & \cr 
         & $E_0+1$ & $\cases{J_0+1 \cr J_0 \cr J_0-1}$ 
      \cr 
      &&\cr 
      \hline 
       &&\cr 
          & $E_0+\ft52$ & $\cases{J_0+1 \cr J_0 \cr J_0-1}$ 
    \cr 
       &&\cr 
      $1$ & $E_0+\ft32$ & 
      $\cases{J_0+2\cr J_0+1\cr J_0+1\cr J_0\cr J_0\cr J_0\cr J_0-1\cr J_0-1\cr J_0-2}$ 
    \cr 
       &&\cr 
          & $E_0+\ft12$ & 
      $\cases{J_0+1\cr J_0\cr J_0-1}$ \cr 
       &&\cr 
      \hline 
       &&\cr 
         & $E_0+3$ & $\cases{J_0}$ \cr 
       &&\cr 
         & $E_0+2$ & 
      $\cases{J_0+2\cr J_0+1\cr J_0+1\cr J_0\cr J_0\cr J_0\cr J_0-1\cr J_0-1\cr J_0-2}$ 
    \cr 
      $\ft12$ &&\cr 
       &&\cr 
         & $E_0+1$ & 
      $\cases{J_0+2\cr J_0+1\cr J_0+1\cr J_0\cr J_0\cr J_0\cr J_0-1\cr J_0-1\cr J_0-2}$ 
    \cr 
       &&\cr 
         & $E_0$ & $\cases{J_0}$ \cr 
      \hline 
       &&\cr 
          & $E_0+\ft52$ & 
      $\cases{J_0+1\cr J_0\cr J_0-1}$ \cr 
       &&\cr 
      $0$ & $E_0+\ft32$ & 
      $\cases{J_0+2\cr J_0+1\cr J_0+1\cr J_0 \cr J_0 \cr J_0-1\cr J_0-1\cr J_0-2}$ 
    \cr 
       &&\cr 
          & $E_0+\ft12$ & 
      $\cases{J_0+1\cr J_0\cr J_0-1}$\cr 
       &&\cr 
      \hline 
    \end{tabular} 
\qquad 
    \begin{tabular}{|c|c|c|} 
      \hline 
      \multicolumn{3}{|c|}{} \cr 
      \multicolumn{3}{|c|}{$DS(2,E_0>5/2,1\vert 3)$} \cr 
      \multicolumn{3}{|c|}{} \cr 
      \hline 
      spin & energy & isospin  \cr 
      \hline 
      \hline 
       &&\cr 
      $2$ & $E_0+\ft32$ & $1$ \cr 
       &&\cr 
      \hline 
       &&\cr 
        & $E_0+2$ & $\cases{2 \cr 1 \cr 0}$ 
      \cr 
      $\ft32$ & & \cr 
         & $E_0+1$ & $\cases{2 \cr 1 \cr 0}$ 
      \cr 
       &&\cr 
      \hline 
       &&\cr 
          & $E_0+\ft52$ & $\cases{2 \cr 1 \cr 0}$ 
    \cr 
       &&\cr 
      $1$ & $E_0+\ft32$ & 
      $\cases{3\cr 2\cr 2\cr 1\cr 1\cr 1 \cr 0}$ 
    \cr 
       &&\cr 
          & $E_0+\ft12$ & 
      $\cases{2\cr 1 \cr 0}$ \cr 
       &&\cr 
      \hline 
       &&\cr 
         & $E_0+3$ & $\cases{1}$ \cr 
       &&\cr 
         & $E_0+2$ & 
       $\cases{3\cr 2\cr 2\cr 1\cr 1\cr 1 \cr 0}$ 
    \cr 
      $\ft12$ &&\cr 
         & $E_0+1$ & 
       $\cases{3\cr 2\cr 2\cr 1\cr 1\cr 1 \cr 0}$ 
    \cr 
       &&\cr 
         & $E_0$ & $\cases{1}$ \cr 
       &&\cr 
      \hline 
       &&\cr 
          & $E_0+\ft52$ & 
      $\cases{2\cr 1 \cr 0}$ \cr 
       &&\cr 
      $0$ & $E_0+\ft32$ & 
       $\cases{3\cr 2\cr 2\cr 1\cr 1 \cr 0}$ 
    \cr 
       &&\cr 
          & $E_0+\ft12$ & 
      $\cases{2\cr 1\cr 0}$\cr 
       &&\cr 
      \hline 
    \end{tabular} 
\qquad 
    \begin{tabular}{|c|c|c|} 
      \hline 
      \multicolumn{3}{|c|}{} \cr 
      \multicolumn{3}{|c|}{$DS(2,E_0>3/2,0\vert 3)$} \cr 
      \multicolumn{3}{|c|}{} \cr 
      \hline 
      spin & energy & isospin  \cr 
      \hline 
      \hline 
       &&\cr 
      $2$ & $E_0+\ft32$ & $0$ \cr 
       &&\cr 
      \hline 
       &&\cr 
        & $E_0+2$ & $\cases{1}$ 
      \cr 
      $\ft32$ & & \cr 
         & $E_0+1$ & $\cases{1}$ 
      \cr 
       &&\cr 
      \hline 
       &&\cr 
          & $E_0+\ft52$ & $\cases{1}$ 
    \cr 
       &&\cr 
      $1$ & $E_0+\ft32$ & 
      $\cases{2\cr 1\cr 0}$ 
    \cr 
       &&\cr 
          & $E_0+\ft12$ & 
      $\cases{1}$ \cr 
       &&\cr 
      \hline 
       &&\cr 
         & $E_0+3$ & $\cases{0}$ \cr 
       &&\cr 
         & $E_0+2$ & 
       $\cases{2\cr 1 \cr 0}$ 
    \cr 
      $\ft12$ &&\cr 
         & $E_0+1$ & 
       $\cases{2\cr 1 \cr 0}$ 
    \cr 
       &&\cr 
         & $E_0$ & $\cases{0}$ \cr 
       &&\cr 
      \hline 
       &&\cr 
          & $E_0+\ft52$ & 
      $\cases{1}$ \cr 
       &&\cr 
      $0$ & $E_0+\ft32$ & 
       $\cases{2\cr 1}$ 
    \cr 
       &&\cr 
          & $E_0+\ft12$ & 
      $\cases{1}$\cr 
       &&\cr 
      \hline 
    \end{tabular} 
\qquad 
    \caption{The long ${\cal N}=3$ graviton multiplet: $E_0 > J_0 + \ft32$. 
         From left to right: $J_0\geq 2, J_0=1, J_0=0$.} 
    \label{longgraviton} 
  \end{center} 
\end{table} 
\begin{table}[htbp] 
  \size 
  \begin{center} 
    \begin{tabular}{|c|c|c|} 
      \hline 
      \multicolumn{3}{|c|}{} \cr 
      \multicolumn{3}{|c|}{$DS(3/2,E_0>J+1,J\ge 2\vert 3)$} \cr 
      \multicolumn{3}{|c|}{} \cr 
      \hline 
      spin & energy & isospin  \cr 
      \hline 
      \hline 
       &&\cr 
      $\ft32$ & $E_0+\ft32$ & $J_0$ \cr 
       &&\cr 
      \hline 
       &&\cr 
          & $E_0+2$ & $\cases{J_0+1\cr J_0\cr J_0-1}$ \cr 
      $1$ &&\cr 
          & $E_0+1$ & $\cases{J_0+1\cr J_0\cr J_0-1}$ \cr 
       &&\cr 
      \hline 
       &&\cr 
      $\ft12$ & $E_0+\ft52$ & $\cases{J_0+1\cr J_0\cr J_0-1}$ \cr 
       &&\cr 
      $\ft12$ & $E_0+\ft32$ & 
      $\cases{J_0+2\cr J_0+1\cr J_0+1\cr J_0\cr J_0 \cr J_0-1\cr J_0-1\cr J_0-2}$ \cr 
       &&\cr 
      $\ft12$ & $E_0+\ft12$ & $\cases{J_0+1\cr J_0\cr J_0-1}$ \cr 
       &&\cr 
      \hline 
       &&\cr 
         & $E_0+3$ & $\cases{J_0}$ \cr 
       &&\cr 
         & $E_0+2$ & $\cases{J_0+2\cr J_0+1\cr J_0\cr J_0\cr J_0-1\cr J_0-2}$ \cr 
      $0$ &&\cr 
         & $E_0+1$ & $\cases{J_0+2\cr J_0+1\cr J_0\cr J_0\cr J_0-1\cr J_0-2}$ \cr 
       &&\cr 
         & $E_0$ & $\cases{J_0}$ \cr 
       &&\cr 
      \hline 
    \end{tabular} 
\qquad 
    \begin{tabular}{|c|c|c|} 
      \hline 
      \multicolumn{3}{|c|}{} \cr 
      \multicolumn{3}{|c|}{$DS(3/2,E_0>2,1\vert 3)$} \cr 
      \multicolumn{3}{|c|}{} \cr 
      \hline 
      spin & energy & isospin  \cr 
      \hline 
      \hline 
       &&\cr 
      $\ft32$ & $E_0+\ft32$ & $1$ \cr 
       &&\cr 
      \hline 
       &&\cr 
          & $E_0+2$ & $\cases{2\cr 1\cr 0}$ \cr 
      $1$ &&\cr 
          & $E_0+1$ & $\cases{2\cr 1\cr 0}$ \cr 
       &&\cr 
      \hline 
       &&\cr 
      $\ft12$ & $E_0+\ft52$ & $\cases{2\cr 1\cr 0}$ \cr 
       &&\cr 
      $\ft12$ & $E_0+\ft32$ & 
      $\cases{3\cr 2\cr 2\cr 1\cr 1 \cr 0}$ \cr 
       &&\cr 
      $\ft12$ & $E_0+\ft12$ & $\cases{2\cr 1\cr 0}$ \cr 
       &&\cr 
      \hline 
       &&\cr 
         & $E_0+3$ & $\cases{1}$ \cr 
       &&\cr 
         & $E_0+2$ & $\cases{3\cr 2\cr 1\cr 1}$ \cr 
      $0$ &&\cr 
         & $E_0+1$ & $\cases{3\cr 2\cr 1\cr 1}$ \cr 
       &&\cr 
         & $E_0$ & $\cases{1}$ \cr 
       &&\cr 
      \hline 
    \end{tabular} 
\qquad 
    \begin{tabular}{|c|c|c|} 
      \hline 
      \multicolumn{3}{|c|}{} \cr 
      \multicolumn{3}{|c|}{$DS(3/2,E_0>1,0\vert 3)$} \cr 
      \multicolumn{3}{|c|}{} \cr 
      \hline 
      spin & energy & isospin  \cr 
      \hline 
      \hline 
       &&\cr 
      $\ft32$ & $E_0+\ft32$ & $0$ \cr 
       &&\cr 
      \hline 
       &&\cr 
          & $E_0+2$ & $\cases{1}$ \cr 
      $1$ &&\cr 
          & $E_0+1$ & $\cases{1}$ \cr 
       &&\cr 
      \hline 
       &&\cr 
      $\ft12$ & $E_0+\ft52$ & $\cases{1}$ \cr 
       &&\cr 
      $\ft12$ & $E_0+\ft32$ & 
      $\cases{2\cr 1}$ \cr 
       &&\cr 
      $\ft12$ & $E_0+\ft12$ & $\cases{1}$ \cr 
       &&\cr 
      \hline 
       &&\cr 
         & $E_0+3$ & $\cases{0}$ \cr 
       &&\cr 
         & $E_0+2$ & $\cases{2\cr 0}$ \cr 
      $0$ &&\cr 
         & $E_0+1$ & $\cases{2\cr 0}$ \cr 
       &&\cr 
         & $E_0$ & $\cases{0}$ \cr 
       &&\cr 
      \hline 
    \end{tabular} 
    \caption{The long ${\cal N}=3$ gravitino multiplet: $E_0 > J_0 + 1$. 
         From left to right: $J_0\geq 2, J_0=1, J_0=0$.} 
    \label{longgravitino} 
  \end{center} 
\end{table} 
\begin{table}[htbp] 
  \begin{center} 
    \size 
    \begin{tabular}{|c|c|c|} 
      \hline 
      \multicolumn{3}{|c|}{} \cr 
      \multicolumn{3}{|c|}{$DS(2,J+3/2,J\ge 2\vert 3)$} \cr 
      \multicolumn{3}{|c|}{} \cr 
      \hline 
      spin & energy & isospin  \cr 
      \hline 
      \hline 
       &&\cr 
      $2$ & $J_0+3$ & $J_0$ \cr 
       &&\cr 
      \hline 
       &&\cr 
        & $J_0+\ft72$ & $\cases{ J_0 \cr J_0-1}$ 
      \cr 
      $\ft32$ & & \cr 
         & $J_0+\ft52$ & $\cases{J_0+1 \cr J_0 \cr J_0-1}$ 
      \cr 
       &&\cr 
      \hline 
       &&\cr 
          & $J_0+4$ & $\cases{J_0-1}$ 
    \cr 
       &&\cr 
      $1$ & $J_0+3$ & 
      $\cases{J_0+1\cr J_0\cr J_0\cr J_0-1\cr J_0-1\cr J_0-2}$ 
    \cr 
       &&\cr 
          & $J_0+2$ & 
      $\cases{J_0+1\cr J_0\cr J_0-1}$ \cr 
       &&\cr 
      \hline 
       &&\cr 
         & $J_0+\ft72$ & 
      $\cases{ J_0\cr J_0-1\cr J_0-2}$ 
    \cr 
      $\ft12$ &&\cr 
         & $J_0+\ft52$ & 
      $\cases{ J_0+1\cr  J_0\cr J_0\cr J_0-1\cr J_0-1\cr J_0-2}$ 
    \cr 
       &&\cr 
         & $J_0+\ft32$ & $\cases{J_0}$ \cr 
       &&\cr 
      \hline 
       &&\cr 
      $0$ & $J_0+3$ & 
      $\cases{J_0\cr J_0-1\cr J_0-2}$ 
    \cr 
       &&\cr 
          & $J_0+2$ & 
      $\cases{ J_0\cr J_0-1}$\cr 
       &&\cr 
      \hline 
    \end{tabular} 
\qquad 
    \begin{tabular}{|c|c|c|} 
      \hline 
      \multicolumn{3}{|c|}{} \cr 
      \multicolumn{3}{|c|}{$DS(2,5/2,1\vert 3)$} \cr 
      \multicolumn{3}{|c|}{} \cr 
      \hline 
      spin & energy & isospin  \cr 
      \hline 
      \hline 
       &&\cr 
      $2$ & $4$ & $1$ \cr 
       &&\cr 
      \hline 
       &&\cr 
        & $\ft92$ & $\cases{1\cr 0}$ 
      \cr 
      $\ft32$ & & \cr 
         & $\ft72$ & $\cases{2\cr 1\cr 0}$ 
      \cr 
       &&\cr 
      \hline 
       &&\cr 
          & $5$ & $\cases{0}$ 
    \cr 
       &&\cr 
      $1$ & $4$ & 
      $\cases{2\cr 1\cr 1\cr 0}$ 
    \cr 
       &&\cr 
          & $3$ & 
      $\cases{2\cr 1\cr 0}$ \cr 
       &&\cr 
      \hline 
       &&\cr 
         & $\ft92$ & 
      $\cases{1}$ 
    \cr 
      $\ft12$ &&\cr 
         & $\ft72$ & 
      $\cases{2\cr 1\cr 1\cr 0}$ 
    \cr 
       &&\cr 
         & $\ft52$ & $\cases{1}$ \cr 
       &&\cr 
      \hline 
       &&\cr 
      $0$ & $4$ & 
      $\cases{1}$ 
    \cr 
       &&\cr 
          & $3$ & 
      $\cases{1 \cr 0}$\cr 
       &&\cr 
      \hline 
    \end{tabular} 
\qquad 
    \begin{tabular}{|c|c|c|} 
      \hline 
      \multicolumn{3}{|c|}{} \cr 
      \multicolumn{3}{|c|}{$DS(2,3/2,0\vert 3)$} \cr 
      \multicolumn{3}{|c|}{} \cr 
      \hline 
      spin & energy & isospin  \cr 
      \hline 
      \hline 
       &&\cr 
      $2$ & $3$ & $0$ \cr 
       &&\cr 
      \hline 
       &&\cr 
      $\ft32$ 
         & $\ft52$ & $\cases{1}$ 
      \cr 
       &&\cr 
      \hline 
       &&\cr 
      $1$  & $2$ & 
      $\cases{1}$ \cr 
       &&\cr 
      \hline 
       &&\cr 
      $\ft12$  & $\ft32$ & $\cases{0}$ \cr 
       &&\cr 
      \hline 
    \end{tabular} 
\qquad 
    \caption{The short ${\cal N}=3$ graviton multiplet: $E_0 = J_0 + \ft32$. 
         From left to right: $J_0\geq 2, J_0=1,$ and $J_0=0$ 
         (that is massless).} 
    \label{shortgraviton} 
  \end{center} 
\end{table} 
\begin{table}[htbp] 
  \size 
  \begin{center} 
    \begin{tabular}{|c|c|c|} 
      \hline 
      \multicolumn{3}{|c|}{} \cr 
      \multicolumn{3}{|c|}{$DS(3/2,J_0+1,J_0\ge 2\vert 3)$} \cr 
      \multicolumn{3}{|c|}{} \cr 
      \hline 
      spin & energy & isospin  \cr 
      \hline 
      \hline 
       &&\cr 
      $\ft32$ & $J_0+\ft52$ & $J_0$ \cr 
       &&\cr 
      \hline 
       &&\cr 
          & $J_0+3$ & $\cases{J_0\cr J_0-1}$ \cr 
      $1$ &&\cr 
          & $J_0+2$ & $\cases{J_0+1\cr J_0\cr J_0-1}$ \cr 
       &&\cr 
      \hline 
       &&\cr 
      $\ft12$ & $J_0+\ft72$ & $\cases{J_0-1}$ \cr 
       &&\cr 
      $\ft12$ & $J_0+\ft52$ & 
      $\cases{J_0+1\cr J_0\cr J_0 \cr J_0-1\cr J_0-1\cr J_0-2}$ \cr 
       &&\cr 
      $\ft12$ & $J_0+\ft32$ & $\cases{J_0+1\cr J_0\cr J_0-1}$ \cr 
       &&\cr 
      \hline 
       &&\cr 
         & $J_0+3$ & $\cases{J_0\cr J_0-1\cr J_0-2}$ \cr 
      $0$ &&\cr 
         & $J_0+2$ & $\cases{J_0+1\cr J_0\cr J_0\cr J_0-1\cr J_0-2}$ \cr 
       &&\cr 
         & $J_0+1$ & $\cases{J_0}$ \cr 
       &&\cr 
      \hline 
    \end{tabular} 
\qquad 
    \begin{tabular}{|c|c|c|} 
      \hline 
      \multicolumn{3}{|c|}{} \cr 
      \multicolumn{3}{|c|}{$DS(3/2,2,1\vert 3)$} \cr 
      \multicolumn{3}{|c|}{} \cr 
      \hline 
      spin & energy & isospin  \cr 
      \hline 
      \hline 
       &&\cr 
      $\ft32$ & $\ft72$ & $1$ \cr 
       &&\cr 
      \hline 
       &&\cr 
          & $4$ & $\cases{1\cr 0}$ \cr 
      $1$ &&\cr 
          & $3$ & $\cases{2\cr 1\cr 0}$ \cr 
       &&\cr 
      \hline 
       &&\cr 
      $\ft12$ & $\ft92$ & $\cases{0}$ \cr 
       &&\cr 
      $\ft12$ & $\ft72$ & 
      $\cases{2\cr 1\cr 1 \cr 0}$ \cr 
       &&\cr 
      $\ft12$ & $\ft52$ & $\cases{2\cr 1\cr 0}$ \cr 
       &&\cr 
      \hline 
       &&\cr 
         & $4$ & $\cases{1}$ \cr 
      $0$ &&\cr 
         & $3$ & $\cases{2\cr 1\cr 1}$ \cr 
       &&\cr 
         & $2$ & $\cases{1}$ \cr 
       &&\cr 
      \hline 
    \end{tabular} 
\qquad 
    \begin{tabular}{|c|c|c|} 
      \hline 
      \multicolumn{3}{|c|}{} \cr 
      \multicolumn{3}{|c|}{$DS(3/2,1,0\vert 3)$} \cr 
      \multicolumn{3}{|c|}{} \cr 
      \hline 
      spin & energy & isospin  \cr 
      \hline 
      \hline 
       &&\cr 
      $\ft32$ & $\ft52$ & $0$ \cr 
       &&\cr 
      \hline 
       &&\cr 
      $1$ & $2$ & $\cases{1}$ \cr 
       &&\cr 
      \hline 
       &&\cr 
      $\ft12$ & $\ft32$ & $\cases{1}$ \cr 
       &&\cr 
      \hline 
       &&\cr 
         & $2$ & $\cases{0}$ \cr 
    $0$ &  & \cr 
         & $1$ & $\cases{0}$ \cr 
       &&\cr 
      \hline 
    \end{tabular} 
    \caption{The short ${\cal N}=3$ gravitino multiplet: $E_0 = J_0 + \ft32$. 
    From left to right, $J_0\geq 2, J_0=1,$ and $J_0=0$ (that is massless).} 
    \label{shortgravitino} 
  \end{center} 
\end{table} 
\begin{table}[htbp] 
  \size 
  \begin{center} 
    \begin{tabular}{|c|c|c|} 
      \hline 
      \multicolumn{3}{|c|}{} \cr 
      \multicolumn{3}{|c|}{$DS(1,J_0,J_0\ge 2\vert 3)$} \cr 
      \multicolumn{3}{|c|}{} \cr 
      \hline 
      spin & energy & isospin  \cr 
      \hline 
      \hline 
       &&\cr 
      $1$  & $J_0+1$ & $\cases{J_0-1}$ \cr 
       &&\cr 
      \hline 
       &&\cr 
      $\ft12$ & $J_0+\ft32$ & 
      $\cases{J_0-1\cr J_0-2}$ \cr 
       &&\cr 
      $\ft12$ & $J_0+\ft12$ & $\cases{J_0\cr J_0-1}$ \cr 
       &&\cr 
      \hline 
       &&\cr 
         & $J_0+2$ & $\cases{J_0-2}$ \cr 
      $0$ &&\cr 
         & $J_0+1$ & $\cases{J_0\cr J_0-1\cr J_0-2}$ \cr 
       &&\cr 
         & $J_0$ & $\cases{J_0}$ \cr 
       &&\cr 
      \hline 
    \end{tabular} 
\qquad 
    \begin{tabular}{|c|c|c|} 
      \hline 
      \multicolumn{3}{|c|}{} \cr 
      \multicolumn{3}{|c|}{$DS(1,1,1\vert 3)$} \cr 
      \multicolumn{3}{|c|}{} \cr 
      \hline 
      spin & energy & isospin  \cr 
      \hline 
      \hline 
       &&\cr 
      $1$  & $2$ & $\cases{0}$ \cr 
       &&\cr 
      \hline 
       &&\cr 
      $\ft12$ & $\ft32$ & $\cases{1}$ \cr 
       &&\cr 
      \hline 
       &&\cr 
      $0$   & $2$ & $\cases{1}$ \cr 
       &&\cr 
         & $1$ & $\cases{1}$ \cr 
       &&\cr 
      \hline 
    \end{tabular} 
    \caption{The ${\cal N}=3$ vector multiplets: $E_0 = J_0$. The massive 
    vector multiplet with $J_0\geq 2$ and the massless vector multiplet with $J_0=1$.} 
    \label{shortvector} 
  \end{center} 
\end{table} 
\end{document}